\documentclass[aip,cha,reprint]{revtex4-1}
\usepackage{amsmath}
\usepackage{amssymb}
\usepackage{graphicx}
\usepackage{dcolumn}
\usepackage{bm}
\usepackage{natbib}
\usepackage{psfrag}

\def\const{\mbox{const}}
\def\w{\omega}

\begin{document}
\title{Reconstructing phase dynamics of oscillator networks}

\author{Bj\"orn Kralemann}
\affiliation{Institut f\"ur P\"adagogik, Christian-Albrechts-Universit\"at 
zu Kiel\\
Olshausenstr. 75, 24118 Kiel, Germany}
\author{Arkady Pikovsky}
\affiliation{Institute of Physics and Astronomy, University of Potsdam, 
Karl-Liebknecht-Str. 24/25, 14476 Potsdam-Golm, Germany}
\author{Michael Rosenblum}
\affiliation{Institute of Physics and Astronomy, University of Potsdam, 
Karl-Liebknecht-Str. 24/25, 14476 Potsdam-Golm, Germany}

\date{\today}

\begin{abstract}
We generalize our recent approach to reconstruction of phase dynamics
of coupled oscillators from data
[B. Kralemann \textit{et al.}, Phys. Rev. E, 77, 066205 (2008)]
to cover the case of small networks of coupled periodic units. Starting from the multivariate
time series, we first reconstruct genuine phases and then obtain the coupling functions in terms
of these phases. The partial norms of these coupling functions quantify directed
coupling between oscillators. We illustrate the method by different network motifs for three coupled
oscillators and for random networks of five and nine units. 
We also discuss nonlinear effects in coupling.
\end{abstract}

\maketitle

\begin{quotation}
Many natural and technological systems can be described as networks of coupled oscillators.
A typical problem in their analysis is to find dynamical features, e.g., synchronization, transition to chaos, 
etc., in dependence on the properties of oscillators and of the coupling. 
Here we address the inverse problem: how to find the properties of the coupling from the observed 
dynamics of the oscillators. This may be relevant for
many experimental situations, where the equations of underlying dynamics are not known. 
We present here a method which is based on invariant reconstruction of phase dynamics equations from 
multivariate observations (at least one scalar oscillating observable of each oscillator must be available)
and includes several algorithmic steps which are rather easy to implement numerically. 
We illustrate the method by numerical examples of small networks of van der Pol oscillators. 
\end{quotation}

\section{Introduction}
Dynamics of coupled self-sustained oscillators attracts vast interest of researchers, 
both due to a 
variety of nontrivial effects and to numerous applications in physics, chemistry, and life 
sciences. In modeling, one usually focuses on the relations of the dynamical features, 
such as synchronization patterns, to the coupling structure of the underlying network. 
In experimental data analysis, an inverse problem typically arises, where one wants 
to reveal the interactions from the observed dynamics. 
A simple consideration demonstrates that this task is really challenging and highly nontrivial:
indeed, computation of correlations and or synchronization index between the oscillators does 
not solve the problem,
because the interactions are generally non-reciprocal while both
the correlation and the synchronization index are symmetric measures.

There are exist two approaches for the determination of directional coupling. 
The first one exploits various information-theoretical techniques and quantifies an interaction 
with the help of such asymmetric measures as transfer entropy, 
Granger causality, predictability, etc. 
\cite{Schreiber-00,Frenzel-Pompe-07,Ishiguro_et_al-08,Chicharro-Andrzejak-09,Faes_et_al-08}. 
The second approach, developed in our previous publications 
\cite{Rosenblum-Pikovsky-01,Rosenblum_et_al-02}, 
see also \cite{Smirnov-Bezruchko-03},
is based on estimation of phases from oscillatory time series and subsequent reconstruction of 
phase dynamics equations and quantification of coupling by inspecting and quantifying the 
structure of these equations. 
The disadvantage of the second, dynamical, approach in comparison to the information-theoretical 
one is that it is restricted to the case of weakly coupled oscillatory systems.
However, for this widely encountered case the dynamical approach has clear advantages:
it works with phases, which are most sensitive to interaction, and yields description of 
connectivity which admits a simple interpretation.
This consideration is supported by several comparative studies \cite{Smirnov-Andrzejak-05,Smirnov_et_al-07}. 
(Notice that there is an intermediate group of techniques which apply informatical measures to
time series of phases \cite{Rosenblum_et_al-02,Palus-Stefanovska-03,Vejmelka-Palus-08,Bahraminasab_et_al-08}.) 
The dynamical approach has been tested in a physical experiment 
\cite{Bezruchko-Ponomarenko-Pikovsky-Rosenblum-03}
and used for analysis of connectivity 
in a network of two or several oscillators in the context of cardio-respiratory interaction 
\cite{Mrowka_et_al-03}, 
brain activity \cite{Cimponeriu_et_al-03,Schnitzler-Gross-05,Osterhage_et_al-08}, and 
climate dynamics \cite{Mokhov-Smirnov-06}.

Recently, we have essentially improved the dynamical approach by suggesting a 
technique for invariant 
reconstruction of the phase dynamics equations from bivariate time series 
\cite{Kralemann_et_al-07,Kralemann_et_al-08,damoco}. Invariance in this context means that reconstructed
equations do not depend on the observables used (at least for a wide class of observables). 
This technique was tested on numerical examples as well as in physical experiments with  
coupled metronomes \cite{Kralemann_et_al-07,Kralemann_et_al-08} and with electro-chemical 
oscillators \cite{Blaha_et_al-11}.
In this paper we extend the technique to cover the case of small networks of interacting oscillators.
The paper is organized as follows. In Section~\ref{sec:phasedescr} we discuss phase description of 
oscillator network.
In Section~\ref{sec:mrpdd} we describe the basic method; 
it is illustrated in Sections~\ref{sec:3vdp} and \ref{sec:59vdp}.
We summarize and discuss our results in Section~\ref{sec:concl}.

\section{Phase description of oscillator network} 
\label{sec:phasedescr}
We assume that an individual oscillator is described by variables $\mathbf{x}$ which satisfy 
a system of ODEs $\dot{\mathbf{x}}=\mathbf{G}(\mathbf{x})$ and that this equation system possesses
a stable limit cycle. The latter can be parameterized by 
phase $\varphi$ which grows uniformly in time, $\dot\varphi=\w=\const$, where $\w$ is the 
oscillation frequency. 
Notably, this phase parameterization is unique 
(up to trivial shifts) and invariant with respect to variable transformations
\cite{Kuramoto-84,Pikovsky-Rosenblum-Kurths-01}.

A network of $N$ oscillators can be  represented as
\begin{equation}
 \dot{\mathbf{x}}_k=\mathbf{G}_k(\mathbf{x}_k)+\mathbf{H}_k(\mathbf{x}_1,\mathbf{x}_2,\ldots)\;,\quad k=1,\ldots,N\;,
\label{eq:gennet}
\end{equation}
where coupling functions $\mathbf{H}_k$ generally depend on the states of all oscillators. 
If variables $\mathbf{x}_l$ 
are absent in $\mathbf{H}_k$, we say that there is no direct coupling from $l$ to $k$. 
Note that generally the oscillators can be quite different, e.g., even the dimensions of state 
variables  $\mathbf{x}_k$ can differ. Of course, the functions $\mathbf{G}_k$ can be different as well.

Assuming weakness of the coupling, one can reduce network equations Eq.~(\ref{eq:gennet})
to phase equations 
\begin{equation}
 \dot\varphi_k=\omega_k+h_k(\varphi_1,\varphi_2,\ldots)\;,\quad k=1,\ldots,N\;,
\label{eq:pheq}
\end{equation}
using a perturbation technique 
(see~\cite{Kuramoto-84,Pikovsky-Rosenblum-Kurths-01} for details).
Then, in the first approximation, $h_k$ depends explicitly on phase $\varphi_l$
only if $\mathbf{H}_k$ depends explicitly on $\mathbf{x}_l$, i.e. if there exists a direct link 
$l\to k$.  
However, in the higher approximations, the r.h.s. of Eq.~(\ref{eq:pheq}) may contain 
high-order terms which depend not only on phases of directly coupled oscillators, but on the other
phases as well. 
Thus, generally speaking, phase oscillators Eq.~(\ref{eq:pheq}) are all-to-all coupled. 
Nevertheless, as shown in Section~\ref{sec:mrpdd} below, 
we can decompose the coupling functions $h_k$ into several components, and comparing the 
norms of these components, characterize the partial couplings between particular nodes.
In fact, we attribute links only to those nodes for which the
corresponding norms are sufficiently large. 
Reconstructed in this way, the structure of the phase network (\ref{eq:pheq}) 
correctly (although not exactly) represents the original network (\ref{eq:gennet}).

To specify the relation between the original network structure and that of the phase oscillators, 
we introduce the following terminology. If the function
$\mathbf{H}_k$ can be written as $\mathbf{H}_k=\sum_{j\ne k}H_{kj}(\mathbf{x}_k,\mathbf{x}_j)$, 
we call this type of coupling pairwise. Other terms, containing a combination of several 
variables acting on $k$, e.g. 
$H_{kjl}(\mathbf{x}_k,\mathbf{x}_j,\mathbf{x}_l)$, we call cross-coupling terms.  
If the coupling is pairwise and weak, then a perturbative phase reduction yields, in 
the leading order, only pairwise terms like $h_{kj}(\varphi_k,\varphi_j)$ in
 the system~(\ref{eq:pheq}) . 
As discussed above, cross-coupling terms in the
phase system~(\ref{eq:pheq}) which depend on two or more driving phases, appear 
either due to cross-coupling terms in the original system (\ref{eq:gennet}) or, if the latter has 
pairwise couplings only, due to high-order combinational terms which 
appear in the perturbative reduction to phase dynamics. 

We emphasize that validity of the phase equations (\ref{eq:pheq}) is not restricted to the case of 
very small coupling. Indeed, these equations are valid as long as an attracting invariant torus exists 
in the phase space of the full system (\ref{eq:gennet}). 
This torus can also exist for parameters of coupling when a perturbation approach is not valid anymore. 
In this case, Eq.~(\ref{eq:pheq}) cannot be derived from Eq.~(\ref{eq:gennet}) by means of a perturbation 
technique, but can be, however, 
reconstructed numerically from multivariate time series.
Here, of course, the relation between networks (\ref{eq:pheq})  and (\ref{eq:gennet}) is nontrivial; 
in particular, some nodes may appear as connected
in  (\ref{eq:pheq}) while being disconnected in (\ref{eq:gennet}).

The main idea of our approach is to reconstruct the phase dynamics (\ref{eq:pheq}) from multivariate 
time series; here it is assumed that $k$th component of the series represents the output of the $k$th 
oscillator. Before proceeding with the description of the method, we make a remark on the range of its validity.
An important dynamical regime in system (\ref{eq:pheq}) is that of full or partial synchrony, when the dynamics
of the phases reduces to a stable torus of lower dimension (partial synchrony) or to a stable limit 
cycle (if all oscillators are synchronized). 
In case of synchrony our technique fails, because the data points do not cover the original 
$N$-dimensional torus and the coupling functions cannot be reconstructed. Sufficiently strong  
noise could cause deviation of the trajectory from the synchronization manifold and help to infer
the dynamics, but discussion of this case goes beyond the scope of this paper.

\section{Reconstruction of phase dynamics from data}
\label{sec:mrpdd}

The main assumption behind our approach is that we have multivariate 
observations of coupled oscillators, where at least one scalar oscillating time series 
$y_k(t)=y_k(\mathbf{x}_k(t))$ is available for each oscillator.
The first step is to transform this observable into a cyclic observable. 
Typically this is done via construction of a two-dimensional embedding $(y_k,\bar{y}_k)$,
where $\bar{y}$ can be, e.g., the Hilbert transform of $y(t)$, 
see~\cite{Pikovsky-Rosenblum-Kurths-01,Kralemann_et_al-08} for a detailed discussion. 
Alternatively, one can use for $\bar{y}$ the time derivative of $y$.
(In all cases the data usually requires some preprocessing, e.g., filtering).
If the trajectory in the plane $(y_k,\bar{y}_k)$ 
rotates around some center, one can compute a cyclic variable 
$\theta_i(t)$, e.g., by means of the arctan function.
As has been in detail argued in~\cite{Kralemann_et_al-08}, in this way we obtain 
not the genuine phase $\varphi_k$ of the oscillator which enters Eq.~(\ref{eq:pheq})), 
but a cyclic variable, or protophase. 
The latter depends on the particular embedding and on the parameterization 
of rotations, and is therefore not invariant. This can be seen already from the 
analysis of an autonomous oscillator: the described procedure generally yields a protophase 
$\theta(t)$ which rotates not uniformely, but obeys
\[
\dot\theta=f(\theta)\;.
\]
The transformation to the uniformly rotating genuine phase $\varphi$ reads
\[
\varphi(\theta)=\omega\int_0^\theta\frac{\text{d}\theta'}{f(\theta')} \;,
\]
where $\omega=2\pi\left[\int_0^{2\pi}f^{-1}(\theta')\text{d}\theta'\right]^{-1}$ is the 
frequency of oscillations. 
For practical reason it is convenient to determine not the function $f$, 
but the probability density of the protophase $\sigma(\theta)=\frac{\omega}{2\pi f(\theta)}$. 
Thus, the 
transformation from the protophase $\theta$ to the phase $\varphi$ can be written as
\[
\varphi(\theta)=2\pi\int_0^\theta \text{d}\theta'\;\sigma(\theta') \;.
\]
Hence, determination of the genuine phase from the data reduces to a problem of finding 
the probability distribution density of the obtained protophase $\theta$, 
which is a standard task in the statistical data analysis. 
Practically, one can use either a Fourier representation of the density $\sigma(\theta)$,
as in~\cite{Kralemann_et_al-08}, or a kernel function representation of $\sigma(\theta)$. 

After the transformation $\theta_k(t)\to\varphi_k(t)$ is performed, we reconstruct the phase dynamics 
making use of the fact that the time derivatives $\dot\varphi_k$ are $2\pi$-periodic functions of 
the phases, in accordance with Eq.~(\ref{eq:pheq}).
These functions can be obtained from the observed multivariate time series of phases $\Phi_k(t)$
with the help of a spectral representation technique~\cite{Kralemann_et_al-08} or by means
of a kernel function estimation. 
Here we present the formulae for the spectral technique (see 
Sec.~IV-A in~\cite{Kralemann_et_al-08} for details), generalized for the case of more 
than two oscillators:
\begin{equation}
\begin{aligned}
F^{(k)}(\varphi_1,\varphi_2,\ldots)&=\frac{F_u^{(k)}(\varphi_1,\varphi_2,\ldots)}
{F_d^{(k)}(\varphi_1,\varphi_2,\ldots)}\;,\\
F_{u,d}&=\sum_{l_1,l_2,\ldots} f^{(u,d)}_{l_1,l_2,\ldots}\exp(il_1\varphi_1+il_2\varphi_2+\ldots)\;,\\
f^{(u)}_{l_1,l_2,\ldots}&=\frac{1}{T}\int_0^{\Phi_k(T)} \text{d}\Phi_k\, \exp(il_1\Phi_1+il_2\Phi_2+\ldots)\;,\\
f^{(d)}_{l_1,l_2,\ldots}&=\frac{1}{T}\int_0^T \text{d}t\, \exp(il_1\Phi_1+il_2\Phi_2+\ldots)\;.
\end{aligned}
\label{eq:spmeth}
\end{equation}
As a result, we obtain the system of the phase dynamics equations in the form
\begin{equation}
\frac{d\varphi_k}{dt}=F^{(k)}(\varphi_1,\varphi_2,\ldots)=\sum_{l_1,l_2,\ldots}
\mathcal{F}^{(k)}_{l_1,l_2,\ldots}\exp(il_1\varphi_1+il_2\varphi_2+\ldots)\;.
\label{eq:pdeq}
\end{equation}

The coupling functions $F^{(k)}$ describe all types of interactions: pairwise, triple-wise, quadruple-wise, 
and so on.
To characterize them separately, we calculate the norms of different coupling 
terms (the partial norms), as follows. 
The term $\mathcal{F}^{k}_{0,0,\ldots}$ is a constant (phase-independent) one, 
it corresponds to the natural frequency of oscillations. 
The pairwise action of oscillator $j$ on oscillator $k$ is determined by those components of $F^{(k)}$ 
which depend on phases $\varphi_k$ and
 $\varphi_j$ only. 
We quantify this action by the partial norm ${\cal N}^{(2)}_{k|j}$; note that here the upper index 
corresponds to the order of interaction (pairwise here).
The partial norm can be computed as
\begin{equation}
\left[{\cal N}^{(2)}_{k|j}\right]^2=
\sum_{l_k;l_j\neq 0}\left|\mathcal{F}^{(k)}_{0,\ldots,0,l_k,0,\ldots,0,l_j,0,\ldots)}\right|^2\;.
\label{eq:2norm}
\end{equation}
Correspondingly, the joint action of oscillators $j,m$ on oscillator $k$ is determined by the cross-coupling
terms containing three phases $\varphi_k,\varphi_j,\varphi_m$. This action is quantified by the 
following partial norm:
\begin{equation}
\left[{\cal N}^{(3)}_{k|jm}\right]^2=\sum_{l_k;l_j\neq 0;l_m\neq 0}\left|
\mathcal{F}^{(k)}_{0,\ldots,0,l_k,0,\ldots,0,l_j,0,\ldots,0,l_m,0,\ldots)}\right|^2\;.
\label{eq:3norm}
\end{equation}
Similarly we can compute the partial norm ${\cal N}^{(4)}_{k|jmn}$, and so on.

Now we discuss the terms $\mathcal{F}^{(k)}_{0,\ldots,0,l_k,0,\ldots,0}$ which depend only on one phase 
$\varphi_k$ and therefore cannot be attributed to any interaction. 
In Ref.~\cite{Kralemann_et_al-08} we used an additional 
transformation $\varphi\to\varphi'$ which 
eliminated these terms, using some additional assumptions on
the structure of coupling. 
In examples presented below we checked that the difference between the
formulations in terms of $\varphi$ and $\varphi'$ is rather small.
Therefore we use the technique described above and neglect the small terms
$\mathcal{F}^{(k)}_{0,\ldots,0,l_k,0,\ldots,0}$.

\section{Case study: Three coupled van der Pol oscillators}
\label{sec:3vdp}

\begin{figure}
\centering
\includegraphics[width=0.6\columnwidth]{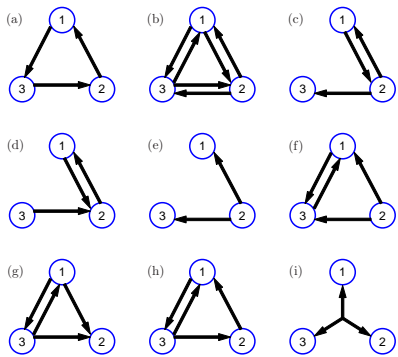}
\caption{
Different configurations in a network of 3 oscillators. (a)-(h): pairwise interaction.
Panel (i) illustrates cross-coupling: an arrow from the center to, say, first oscillator reflects the term
$\sim x_2x_3$ with $\sigma_{11}=1$ in Eq.~(\ref{eq:3vdp}), etc.
}
\label{fig:cofigs}
\end{figure}

In this section we test the presented technique on a system of three coupled van der Pol oscillators:
\begin{equation}
\begin{aligned}
 \ddot x_1-\mu(1-x_1^2)\dot x_1+\w_1^2 x_1&=\Delta\left [\sigma_{12}(\dot x_2+x_2)+
                  \sigma_{13}(\dot x_3+x_3)\right ]\\ 
               &+\sigma_{11}\eta x_2x_3\;,\\
 \ddot x_2-\mu(1-x_2^2)\dot x_2+\w_2^2 x_2&=\Delta\left [\sigma_{21}(\dot x_1+x_1)+
                   \sigma_{23}(\dot x_3+x_3)\right ]\\&+\sigma_{22}\eta x_1x_3\;,\\
 \ddot x_3-\mu(1-x_3^2)\dot x_3+\w_3^2 x_3&=\Delta\left [\sigma_{31}(\dot x_1+x_1)+
                   \sigma_{32}(\dot x_2+x_2)\right ]\\&+\sigma_{33}\eta x_1x_2\;.
\end{aligned}
\label{eq:3vdp}
\end{equation}
Here the matrix $\sigma_{kl}$, with entries zero and one, determines the coupling structure 
(topology) of the network: 
for non-diagonal terms, $\sigma_{kl}=1$, if there 
is forcing from oscillator $l$ to oscillator $k$, and $\sigma_{kl}=0$, otherwise. 
Diagonal terms $\sigma_{kk}$ determine the presence or absence of cross-coupling.
Parameters $\Delta$  and $\eta$ describe the intensity of pairwise and cross-coupling, 
respectively. 
In all examples of this section we use the values of parameters 
$\mu=0.5,\;\w_1=1,\;\w_2=1.3247,\;\w_3=1.75483$. 
Equations (\ref{eq:3vdp}) have been solved by the Runge-Kutta method, 
and the time series of $x_k,\dot x_k$ 
were used to calculate the protophases as $\theta_k=-\text{arctan}(\dot x_k,x_k)$. 
The data sets consisted of $10^6$ or $10^7$ points sampled with $0.01$ time step.
The analyzed coupling configurations are schematically given in 
Fig.~\ref{fig:cofigs}. Note that we chose the oscillator frequencies and parameters 
of coupling so that network remains asynchronous.

\begin{figure}
\centering
\includegraphics[width=0.7\columnwidth]{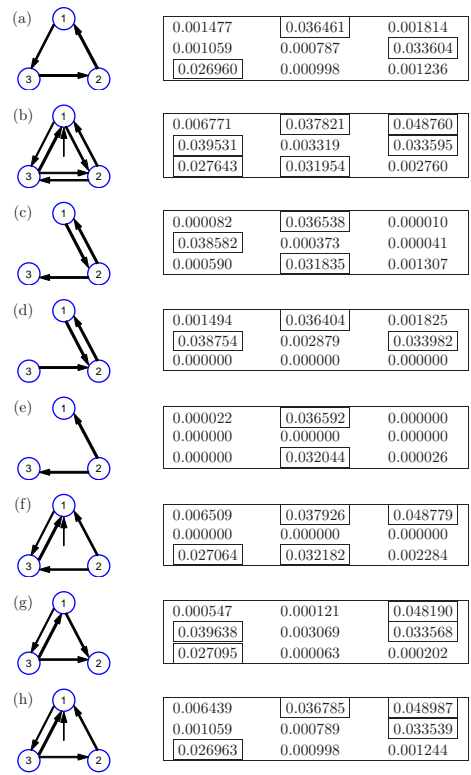}
\caption{Reconsructed network configurations, to be compared with the true configurations, shown 
in Fig.~\ref{fig:cofigs}(a-h). Each panel shows the partial norms of reconstructed coupling functions
as a table; the numbers in boxes correspond to truly existing links, so ideally the numbers without 
the boxes should be zero. Pictures present the largest 
norms only (see text); the values of the norms are coded by width of linking arrows. 
The arrows from the center to the first oscillator (see panels (b), (f), and (h)) reflect
the joint action from oscillators 2 and 3 which cannot be decomposed into pairwise actions 
$2\to 1$ and $3\to 1$. Coupling strength $\Delta=0.05$ and $\eta=0$. 
}
\label{fig:tab1}
\end{figure}

\begin{figure}
\centering
\includegraphics[width=0.7\columnwidth]{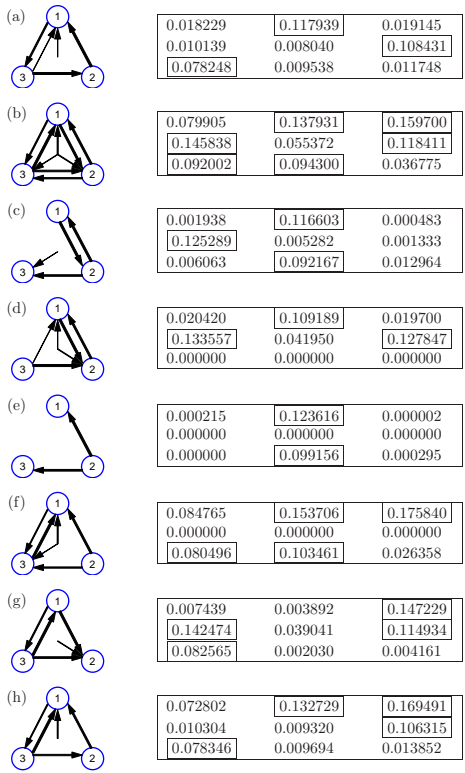}
\caption{Same as Fig.~\ref{fig:tab1}, but for $\Delta=0.15$.
}
\label{fig:tab2}
\end{figure}

In the first set of tests we considered only pairwise coupling in Eqs.~(\ref{eq:3vdp}), 
i.e. we took $\sigma_{kk}=0$. 
The results for $\Delta=0.05$ and $\Delta=0.15$ are presented in 
Figs.~\ref{fig:tab1},\ref{fig:tab2}, respectively.
Here in the left column we schematically show the reconstructed coupling configuration.
The corresponding partial norms are coded by the width of arrows which link
 the nodes.
Only norms which are larger than $10\%$ of the maximal norm are shown here. 
The numerical values of all reconstructed norms can 
be seen in the tables in the right column.
Here we show the coupling norms for the 
pairwise coupling (\ref{eq:2norm}) as non-diagonal terms, and the triple-phase 
coupling according to (\ref{eq:3norm}) as diagonal ones.
 The coupling terms, which are present in the original system
(\ref{eq:3vdp}) (i.e. those with $\sigma_{kl}=1$) are shown in boxes. 
We see that in all cases these terms are 
definitely larger than other entries
of the reconstructed coupling matrix, i.e. the reconstruction works pretty well in all cases. 
Comparing the results for smaller and larger coupling (Fig.~\ref{fig:tab1} and \ref{fig:tab2}, 
respectively), we see that larger coupling strength
leads to appearance of cross-coupling terms for phases that are not present 
in the original system (\ref{eq:3vdp}).

\begin{figure}
\centering
\includegraphics[width=0.7\columnwidth]{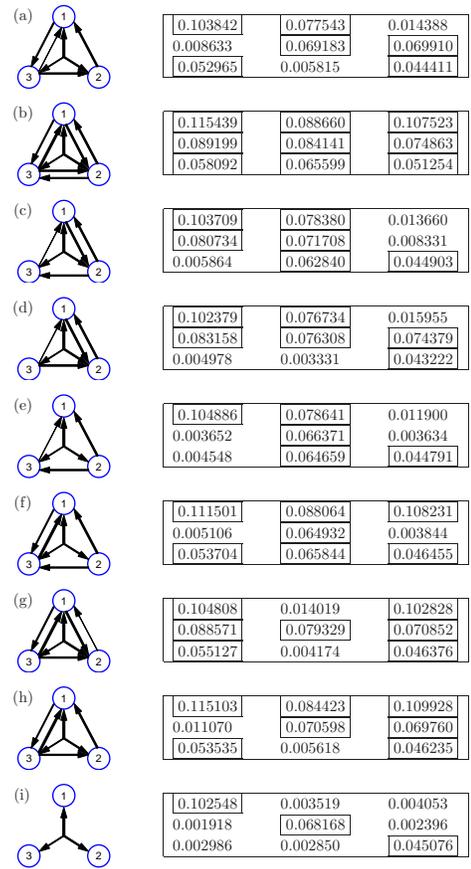}
\caption{Same as Fig.~\ref{fig:tab1}, but for combination of pairwise and triple-coupling, 
$\Delta=0.1$, $\eta=0.1$. Panels (a-h) correspond to combination of pairwise coupling  as 
shown in Fig.~\ref{fig:cofigs}(a-h) with the triple-coupling as in Fig.~\ref{fig:cofigs}(i).
Panel (i) shows the results for purely triple-coupling.
}
\label{fig:tab3}
\end{figure}

In the second set of tests we included the cross-couplings by setting $\sigma_{kk}=1$ and
$\eta=0.1$. 
So, each configuration shown in Fig.~\ref{fig:cofigs}(a-h) was complimented by the links 
according to Fig.~\ref{fig:cofigs}(i). 
The results are presented in Fig.~\ref{fig:tab3}. We see that basically the coupling structure
is correctly reproduced by the method, although due to a joint action
of different terms some couplings, not presented in the original system (\ref{eq:3vdp}), 
do not fall below $10\%$ of the largest norm.

Summarizing the results of this section, we conclude that the connectivity of a 
weakly and pairwisely coupled network can be correctly revealed. 
For stronger coupling, all really existing in the original network links are 
correctly revealed and some additional links appear. 
However, the latter are not necessarily spurios. Indeed, as discussed 
above, not very weak pairwise coupling in the original network may lead to additional 
connections in the network of phase oscillators and to cross-coupling terms.
Our results are in full agreement with this argument.

\section{Case study: Networks of five and nine oscillators}
\label{sec:59vdp}

In this section we report on the results for small networks of $N=5$ and $N=9$ coupled
van der Pol oscillators. 
We simulated the systems with pairwise interaction only:
\begin{equation}
\ddot x_k-\mu(1-x_k^2)\dot x_k+\omega_i^2 x_k=\Delta\sum_{l\neq k} 
            \sigma_{kl}(x_l\cos\alpha_{kl}+\dot x_l\sin\alpha_{kl})\;.
\label{eq:manyvdp}
\end{equation}
As above, $\mu=0.5$, $\Delta$ is intensity of coupling, and $\sigma_{kl}$
is the $N\times N$ connectivity matrix ($\sigma_{kl}=1$ if oscillator 
$l$ acts on oscillator $k$, and $\sigma_{kl}=0$ otherwise). 
An additional parameter $\alpha_{kl}$ describes the phase shift in 
the coupling, so that generally the latter can be both attracting or repelling. 

We note that computational efforts and data requirements for reconstruction of a 
network of $N$ oscillators grow rapidly with $N$, so that full reconstruction for  
$N>3$, though theoretically possible, see Eqs.~(\ref{eq:spmeth},\ref{eq:pdeq}), 
becomes practically unfeasible. On the other hand, in the previous section we have 
verified that weak pairwise couplings can be reliably recovered. 
Therefore, for networks described by Eqs.~(\ref{eq:manyvdp}) we 
reconstruct pairwise coupling functions only, for all pairs of network units.

We performed a statistical analysis of the model (\ref{eq:manyvdp}): 
for many randomly chosen values of frequencies $\omega_k$, parameters 
$\alpha_{kl}$, and connectivity matrices $\sigma_{kl}$ we analysed pairwise 
interactions in the networks.
For $N=5$, the frequencies $\omega_k$  have been chosen uniformly distributed in $(0.5,1.5)$, and the 
number of incoming links for each node was 2 (i.e. $\sum_l\sigma_{kl}=2$), while the links have been 
chosen randomly (the incoming and outgoing links were chosen independently).
For $N=9$, the frequencies $\omega$ were distributed uniformly in $(0.5,2.5)$, and the number 
of incoming links was 3. The intensity of coupling in both cases was $\Delta=0.15$, the phase 
shift $\alpha_{kl}$ was distributed uniformly in $[0,2\pi)$.

For each network we computed $N$ protophases according to
$\theta_k=-\text{arctan}(\dot x_k,\omega_k x_k)$ and then transformed them to phases $\varphi_k$. 
Next,  synchronization analysis have been performed: for all pairs we calculated
the synchronization index $|\langle e^{i(\varphi_k-\varphi_l)}\rangle|$. 
Since our technique does not work in case of synchrony, 
we excluded from the further consideration 
all networks where synchronization index was larger than $0.5$ at least for one link.

\begin{figure}
\centering
\includegraphics[width=0.8\columnwidth]{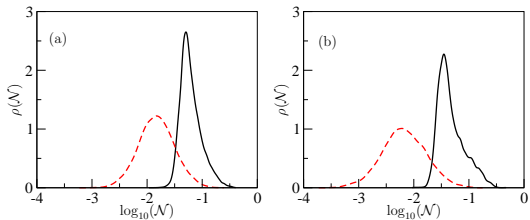}
\caption{Probability distribution densities $\rho({\cal N})$ of the norms of 
reconstructed couplings: black solid line: 
really present connections ($\sigma_{kl}=1$), dashed red line:
absent connections ($\sigma_{kl}=0$), for networks (\ref{eq:manyvdp}) of 5 (a) 
and 9 (b) oscillators.
}
\label{fig:59vdp}
\end{figure}

For each obtained non-synchronous network we fitted the time derivatives $\dot\varphi_k$ by  
$\sum_{k\ne l} F_{kl}(\varphi_k,\varphi_l)$, where $F_{kl}$ depend on two phases only.
Next, we computed norms ${\cal N}_{kl}$ of all $F_{kl}$; 
these norms represent reconstructed connectivity of the network. Performing the
analysis with a large ensemble of non-synchronous networks, we separated ${\cal N}_{kl}$
into two classes: first class contains those values of ${\cal N}_{kl}$ for which $\sigma_{kl}=1$, i.e. 
the connections between nodes $k,l$ in the network (\ref{eq:manyvdp}) really exists. Otherwise, if 
$\sigma_{kl}=0$, the value ${\cal N}_{kl}$ belongs to the second class.
The obtained results are summarized in Fig.~\ref{fig:59vdp}.
Ideally, one would expect that all norms in the first class are much larger than those 
in the second one (cf. Figs.~\ref{fig:tab1},\ref{fig:tab2}). 
We see that there is a clear, although not ideal, separation between 
connected and non-connected links, what allows us to conclude that generally our technique correctly 
reproduces the network structure. 
However, a more detailed statistical analysis is needed before a 
``blind'' application of the method  would be possible.

\section{Discussion and conclusion}
\label{sec:concl}

In summary, we have presented a technique for reconstruction of the phase model 
of a network of coupled limit cycle oscillators. 
From the reconstructed model we infer directional couplings of the network.
We have tested the technique on small networks of  
van der Pol oscillators with pairwise and cross-coupling. We have shown, that presence and direction of 
existing links can be reliably revealed. However, if the coupling is not very weak, 
the technique yields additional links which are absent in the original network.
The differentiation between the truly existing and additional links is not perfect.
Now we discuss  where the additional links come from.
First, some additional links may be numerical artefacts (spurious links). 
They may appear due to noise, closeness to sychrony, neglection of the 
amplitude information, etc. Second, some links, absent in the model in terms 
of state variables, should indeed be present in the phase model and are, 
therefore, not spurious.
If we asume that a matrix $\sigma_{kl}$  determines the structural connectivity 
of the original network model (\ref{eq:gennet}), then the phase model (\ref{eq:pheq}) for 
this network would have an effective 
connectivity matrix $\Sigma_{kl}$ with a larger number of links and possibly with many cross-couplings. 
It is this effective connectivity that is determined in our approach, not the original one.  
Moreover, here one should distinguish between the
structural connectivity (described by matrices $\sigma,\Sigma$) and the
functional one, which is described by similarities in the dynamics (e.g. via correlations or synchronization). 
The functional connectivity (this term, typically used in the context of analysis of brain activity, is 
widely understood as a correlated time behavior) results from the dynamics, and may only loosely correspond 
to the structural one \cite{Tass_et_al-98,Shelter_et_al-06}. 

Finally, we discuss the perspectives of the analysis of networks of many oscillators. 
As we have demonstrated, generally a 
pairwise analysis yields good results only if the connections are pairwise and weak.
In the case of cross-coupling (see e.g., Fig.~\ref{fig:cofigs}(i)), pairwise 
analysis does not help. Determination of applicability of a pairwise analysis 
for a given network remains an open problem. A possible solution might be 
reconstruction of phase dynamics for several or all triplets of oscillators and 
comparison of the terms, dependent on three phases, with those dependent only on two.
If the triple terms are much smaller in norm than pairwise ones, than the pairwise analysis 
may be sufficient.

\begin{acknowledgments}
The research was supported by the Merz-Stiftung, Berlin.
\end{acknowledgments}

%

\end{document}